\documentstyle[aps, preprint, eqsecnum]{revtex}
                                    
\newcommand{\br}{{\bf r}}
\newcommand{\ba}{{\bf a}}
\newcommand{\bk}{{\bf k}}
\newcommand{\dd}{{\cal D}}

\begin{document}
\draft

\preprint{SCU-TP-97-1001; \hskip0.5cm
SNUTP-97-041 }
%GAC%%hep-th/9705059}

\title{Radiation Damping of a BPS Monopole; an Implication to 
S-duality}

\author{ Dongsu Bak\footnote{email address: dsbak@mach.scu.ac.kr} 
and Hyunsoo Min\footnote{email address: hsmin@dirac.scu.ac.kr}}

\address{ Department of Physics,
Seoul City University,
Seoul 130-743, Korea}

\date{\today}

\maketitle
\widetext

\begin{abstract}

The radiation reaction of a BPS monopole
in the presence of incident  electromagnetic waves as well as 
massless Higgs waves is analyzed classically. 
The reactive forces are compared
to those of  $W$ boson that is interpreted as a dual partner 
of the 
BPS monopole. It is shown that the damping of acceleration is 
dual to each other, while in the case of finite size effects 
the duality is broken explicitly. Their implications on the 
duality
are discussed.
\end{abstract}


\section{Introduction}

It is a fascination  that non-singular magnetic monopoles  
arise  
as classical soliton solutions in  certain spontaneously broken 
Yang-Mills gauge theories\cite{thooft,polyakov}. These 
monopoles are 
extended objects with definite mass and couple 
effectively in low energies to   the electromagnetic 
fields\cite{bak1}.
Of particular interest is the BPS limit of the Yang-Mills-Higgs
theory\cite{prasad}  where one may find   static multi-monopole 
solutions
by solving the first order Bogomol'nyi 
equation\cite{bogomolnyi}. 
Dynamics of the
BPS monopole has received a wide  attention recently in 
relation to S-duality 
(electric and magnetic duality)  in supersymmetric 
extensions of the theory, which relates a magnetically weak 
coupling states of 
monopoles to  their electric counter 
parts\cite{montonen,sen,seiberg,vafa,ferrari}. 
In N=4 supersymmetric Yang-Mills theory, this S-duality 
conjecture\cite{montonen,sen} turns
out to hold exactly for the BPS saturated states\cite{ferrari}.

The tests so far are limited to nondynamical BPS saturated 
states and 
their modular-space dynamics at large 
separations\cite{ferrari,manton,atiyah,samols}. 
Observing that even the classical 
dynamics at  weak coupling limits of electric or magnetic 
charges,
comprises nontrivial dynamical excitations from the BPS 
saturated states,
one may further test the S-duality on 
these excited states. 
Especially, the BPS monopole possesses the finite size inversely 
proportional to the mass of the W boson, whereas the W boson 
seems
 pointlike in the classical dynamics.

Thus one inquires  whether the size effect of the BPS monopole 
enters
explicitly  in its detailed classical description. 
Specifically, we 
consider  the responses of the BPS monopole to 
incident electromagnetic waves. The BPS monopole is expected to
undergo a periodic motion and   emits
radiations owing to  its coupling
to the electromagnetic fields and  the massless Higgs 
fields. These
radiations are not included in modular space description 
simply because 
the massless fields are truncated in this approximation. 
In Ref.~\cite{bak1}, 
it was found that ignoring the radiation reaction, the  
duality turns out 
to hold even in the presence of the radiations.

In this note, we focus on the radiation reaction of the BPS 
monopole.  
In the sense that classical motions of the 
BPS monopole are completely fixed by the field equations with an
asymptotic boundary condition, the problem of radiation reaction 
 is perfectly well posed and self-contained. This is contrasted 
to 
the case of the electrodynamics, where the Abraham-Lorentz 
model\cite{abraham} 
or the other attempts\cite{butler} 
in explaining the damping effect,  
are plagued with unnatural assumptions, and  not 
sufficient in themselves\cite{jackson}. 
 We shall examine the radiation 
damping of the BPS monopole and  compare the resulting 
expressions
 to those of the W boson.(W bosons being pointlike, we shall, 
anyway,  follow 
the Abraham-Lorentz scheme in order to obtain reaction effects.) 
The finite size effect of the BPS monopole 
enters the description in consideration of higher-order 
corrections,
and, in the following, it will be analyzed and 
compared to its dual part again.

In Section II, we briefly review the SU(2) Yang-Mills-Higgs 
theory in 
the BPS limit and describe the BPS monopole solutions. In 
addition, we
present an equation describing 
light and Higgs-scalar scattering off a BPS monopole and its 
solution 
to the lowest order. Scattering cross-sections obtaind from 
this solution 
is none other than the dual Thomson formula, confirming the 
duality
to this order.

In section III, we first review the previous attempts to explain 
the radiation damping in electromagnetism. We shall obtain 
the higher oder solutions to the field equation  and give a 
description on
the radiation damping and finite size effect of the BPS 
monopoles.

Last section comprizes the radiation reactions  of the W bosons 
and comparisons to those of the BPS monopole. Conclusions and 
some comments are
followed.

\section{BPS monopoles and scattering by light or  Higgs-scalar}

The simplest field theory possessing magnetic monopoles, is the
SU(2) Yang-Mills-Higgs system with the Higgs fields 
in the adjoint representation. We shall consider this theory 
in a Prasad-Sommerfield limit where the supersymmetric 
extension is natural. The system is described by the 
Lagrangian density $(a = 1,2,3)$
\begin{equation}
\label{lag}
{\cal L} =-{1\over 4} G_a^{\mu\nu}G^a_{\mu\nu}-
{1\over 2} (D_\mu\phi)_a (D^\mu\phi)_a 
\end{equation}
where
\begin{eqnarray}
&&G_a^{\mu\nu} =\partial^\mu A^\nu_a-\partial^\nu A^\mu_a
+e\epsilon_{abc}A^\mu_b A^\nu_c,\\
&&(D_\mu \phi)_a=\partial_\mu\phi_a+e\epsilon_{abc} 
A_{\mu}^b\phi^c,
\end{eqnarray}
and the Higgs fields are subject to the asymptotic boundary
condition
\begin{equation}
\label{bc1}
\phi_a \phi_a \rightarrow f^2 (\neq 0)\ \ \ {\rm as}\  r 
\rightarrow \infty .
\end{equation}
Our metric convention is with signature $(-,+,+,+)$. The field 
equations read
\begin{eqnarray}
\label{fieldeq}
&&(D_\mu G^{\mu\nu})_a =-e\epsilon_{abc}(D^\nu\phi)^b \phi^c,\\
\label{pfieldeq}
&&(D_\mu D^\mu \phi)_a=0 .
\end{eqnarray}
There exist static monopole solutions satisfying the Bogomol'nyi
equations 
\begin{equation}
\label{bogomol}
 B^a_i=\mp (D_i\phi)^a, 
 \ \ \ \ (B^a_i\equiv {1\over 2} \epsilon_{ijk} G_a^{jk})
\end{equation}
and   $A_0^a=0$. Within the spherical symmetric ansatz, a 
unique solution 
to (\ref{fieldeq})-(\ref{pfieldeq}) is\cite{prasad}
\begin{eqnarray}
\label{monopole}
&&A^i_a( {\bf r}) =\epsilon_{aij}{\hat r_j\over er}(1-{m r\over 
\sinh m r}),\\
\label{amonopole}
&& \phi_a ({\bf r})= \pm \hat r_a f (\coth m r -{1\over m r})
\end{eqnarray}
where $m(\equiv ef)$ is the mass of W bosons.

To define charges,
first we introduce relevant asymptotic (i.e. $r\rightarrow 
\infty$) fields by 
\begin{equation}
\label{asymp}
F_{\mu\nu}^{\rm em} = \ { G}^a_{\mu\nu} {\phi^a\over |\bf\phi|},
\ \ \  H_\mu =  - (D_\mu\phi)^a  {\phi^a\over |\bf\phi|},
\end{equation}
where $F_{\mu\nu}^{\rm em}$ and $H_\mu$   describe respectively 
the electromagnetic fields
and the massless Higgs.
The magnetic  and the scalar charges are
defined in a conventional way by the fluxes
\begin{eqnarray}
\label{charge}
&&g =\lim_{r\rightarrow \infty}\int dS^i B^i_{\rm em},\\
&&q_s  = - \lim_{r\rightarrow \infty}\int dS^i H^i.
\end{eqnarray}
Hence the above is the BPS monopole solution with the magnetic 
charge, 
$g= \mp {4\pi/ e}$,
the scalar charge $q_s= {4\pi/e}$ and the mass 
(defined by evaluating the Hamiltonian) 
$M={4\pi f/e}$. Static multi-monopole solutions were also 
constructed by solving
the Bogomol'nyi equation\cite{bogomolnyi}. This is possible 
because the magnetic forces between each monopole are balanced 
by the scalar 
forces  in this BPS limit. 

Note that in the broken phase, there still exists a  massless 
U(1) field that
we interpret as an electromagnetic field. The other two vector 
bosons become 
massive by Higgs mechanism. In the BPS limit, a massless Higgs 
is also 
remained and giving long range attractive interactions between 
the monopoles.

The question how a BPS monopole responds to  asymptotic 
electromagnetic 
waves can be answered by the analysis of the field equations 
since all 
the recipes for the problem exist in the theory. Especially, 
in Ref.~\cite{bak1}, the 
response of the BPS monopole in the presence of an asymptotic 
electromagnetic wave,
specified by
\begin{equation}
\label{emfield}
 {\bf B}^{\rm em}= {M\omega^2 \over g}{\rm Re}
\left[ i \left({\bf a}-({\bf \hat k}\cdot{\bf a}){\bf\hat k} 
\right)\,
e^{i{\bf k}\cdot {\bf x}-i\omega t}\right] \ \ \ \ 
(\omega=|{\bf k}|).
\end{equation}
is considered. The constant real vector $\ba$ in 
(\ref{emfield}) will describe
amplitude and direction of oscillation later.
The solution to the field equations with the above asymptotic 
condition 
is constructed for the case
${\omega/ m}\ll 1$ and  $\omega  a\ll   1$ to the lowest order.

We provide here a brief review on the analysis of the 
scattering problem 
in Ref.~\cite{bak1}.
One begins the analysis by writing the ansatz for the solution
\begin{eqnarray} 
\label{ansatz}
&&A_\mu^a(\br, t) ={\bar A}^a_\mu(\br) + 
{\rm Re}[\alpha^a_\mu (\br)e^{-i\omega t}]+O(a^2),\\
\label{aaansatz}
&&\phi^a(\br, t) ={\bar \phi}^a(\br) + 
{\rm Re}[\pi^a (\br)e^{-i\omega t}]+O(a^2)
\end{eqnarray} 
where  $({\bar A}^a_i,{\bar A}^a_0=0, {\bar\phi}^a )$ is the 
static 
solution in (\ref{monopole})-(\ref{amonopole}) 
and $(\alpha_\mu^a,\pi^a)$ are assumed to be
$O(a)$. The position ${\bf X}$ of the monopole is defined 
by the zero of the field $\phi_a(\br,t)$, which is, of course, a 
gauge invariant quantity. For example,  applying this 
definition, one may say
the static monopole
in (\ref{monopole})-(\ref{amonopole}) 
is located at the origin. Inserting the ansatz into the 
field equations (\ref{fieldeq})-(\ref{pfieldeq}), 
one finds that $(\alpha_\mu,\pi)$ satisfy 
\begin{eqnarray} 
\label{perteq}
&&(D_j G^{ji})_a +e \epsilon_{abc}(D^i\phi)^b\phi^c+
\omega^2\alpha_a^i +i\omega({\bar D}^i\alpha^0)_a=0,\\
\label{aperteq}
&&({\bar D}_k{\bar D}_k\alpha^0)_a-i \omega 
({\bar D}_k\alpha_k)_a
+ie\omega\epsilon_{abc}\pi^b{\bar\phi}^c-e^2 
\epsilon_{abc}\epsilon_{bde}\alpha_0^d{\bar\phi}^e
{\bar\phi}^c=0,\\
\label{bperteq}
&&({ D}_k{D}_k\phi)_a+\omega^2\pi_a
+ie\omega\epsilon_{abc}\alpha^b_0{\bar\phi}^c=0
\end{eqnarray}  
where ${\bar D}_i^{ac}\equiv \partial_i\delta_{ac}+
e\epsilon_{abc}{\bar A}^b_i$ and $O(a^2)$ terms are ignored. 
Once 
the exact solutions of the above equations  are found, all 
the relevant linear effects are included.

These equations are greatly simplified if we introduce 
functions $b^i_a(\br)$
by the relation
\begin{equation}
\label{defb}
 G_a^{ij}(\br, t)= \mp\epsilon_{ijk}[(D_k\phi)^a(\br,t)+
b_k^a(\br)e^{-i\omega t}+O(a^2)].
\end{equation} 
where $b^i_a(\br)$ being $O(a)$.  Using this definition for 
$b^i_a(\br)$, the 
equations of $O(a)$ are reduced to
\begin{eqnarray} 
\label{perteq2}
&&\pi^a={1\over \omega^2}\left[({\bar D}_k b_k)_a -ie\omega 
\epsilon_{abc}\alpha^b_0{\bar\phi}^c\right],\\
\label{aperteq2}
&&\alpha^a_i={1\over \omega^2}\left[\mp\epsilon_{ijk}
({\bar D}_j b_k)^a
+e 
\epsilon_{abc}b^b_i{\bar\phi}^c-i\omega 
({\bar D}_i\alpha_0)^a\right],
\end{eqnarray}  
and
\begin{equation}
\label{helmholtz}
[({\bar D}_k{\bar D}_k+\omega^2) b_i]^a+e^2 
\epsilon_{abc}\epsilon_{bde} b_i^d{\bar\phi}^e{\bar\phi}^c=0.
\end{equation}
Since there is no equation for $\alpha_0$, they are arbitrary 
functions; 
this implies that they are 
in fact  pure gauge degrees of freedom. It is clear that once a 
solution
to (\ref{helmholtz}) is obtained, $(\alpha_i^a,\pi^a)$ are 
automatically given 
by the relations (\ref{perteq2})-(\ref{aperteq2}). 
In Ref.~\cite{bak1}, 
the solution subject to 
the asymptotic condition corresponding to
the incident waves in (\ref{emfield}),
is indeed 
found to the leading order:
\begin{equation}
\label{solutiona}
 b_{i}^a(\br)=\mp i\omega^2 a_i f \coth m r 
e^{i {\bf k}\cdot \br}{\hat r}^a \pm i\omega^2 
a_i{e^{i\omega r}\over er}{\hat r}^a.
\end{equation}

Using  (\ref{aaansatz}), (\ref{perteq2}) and the definition
of monopole position, one may easily show that the motion is 
described
by
\begin{equation}
\label{motiona}
 {\bf X}(t)={\rm Re}[i{\bf a}e^{-i\omega t}] + 
O(a\omega)O(w\!/\!m).
\end{equation}
while straightforward computations lead 
to expressions for the asymptotic fields in the scattering 
region:
\begin{eqnarray} 
\label{asympa}
&&(D_0\phi)^a(\br,t)\sim \mp i\omega^2 {\hat r}^a
\left( {\bf a}\cdot{\bf \hat k}f e^{i\bk\cdot\br -iwt}-
{\ba\cdot {\hat\br}\over er}e^{i\omega r -iwt}\right),\\
&&(D_i\phi)^a(\br,t)\sim \pm i\omega^2 {\hat r}^a
\left( {\bf a}\cdot{\bf \hat k}f e^{i\bk\cdot\br -iwt}{\hat k}_i-
{\ba\cdot {\hat\br}\over er}e^{i\omega r -iwt}
{\hat r}_i\right),\\
&&G^{i0}_a(\br,t)\sim i\omega^2 {\hat r}^a
\left[ ({\bf \hat k}\times {\bf a})_i f e^{i\bk\cdot\br -iwt}-
{({\hat\br}\times \ba)_i}{e^{i\omega r -iwt}\over er}\right],\\
&&G^{ij}_a(\br,t)\sim -i\epsilon^{ijk}\omega^2 {\hat r}^a
\left[({\bf \hat k}\times({\bf \hat k}\times {\bf a}))_k f 
e^{i\bk\cdot\br -iwt}-
{({\hat\br}\times({\hat\br}\times \ba))_k}
{e^{i\omega r -iwt}\over er}\right].
\end{eqnarray}  
The form of incident waves are clear in the above expressions, 
so the 
generalized Lorentz force law can be checked explicitly 
to the order $O(a\omega)$: 
\begin{equation}
\label{lorentza}
 M{\bf {\ddot X}}(t)=[g {\bf B}^{\rm em}_{\rm inc}+q_s 
{\bf H}_{\rm inc}]_{\br={\bf X}} + O(a\omega)O(w\!/\!m),
\end{equation}
where the subscript `inc' indicates that the related quantities
 belong to the incident parts of the fields. From the radiation 
fields---the 
terms $O(r^{-1})$---the related differential 
crossections for the electromagnetic and Higgs waves are 
determined as 
\begin{eqnarray} 
\label{crossection}
&&\left({d\sigma\over d\Omega}\right)_{{\rm em}\rightarrow 
{\rm em}}=
\left({g^2\over 4\pi M}\right)^2 \sin^2\Theta,\\
\label{acrossection}
&& \left({d\sigma\over d\Omega}\right)_{{\rm em}\rightarrow 
{\rm Higgs}}=
\left({g^2\over 4\pi M}\right)\left({q_s^2\over 4\pi M}\right) 
\cos^2\Theta\\
\label{bcrossection}
&&\left({d\sigma\over d\Omega}\right)_{{\rm Higgs}\rightarrow 
{\rm em}}=
\left({q_s^2\over 4\pi M}\right)\left({g^2\over 4\pi M}\right) 
\sin^2\theta, \\
\label{ccrossection}
&& \left({d\sigma\over d\Omega}\right)_{{\rm Higgs}\rightarrow 
{\rm Higgs}}=
\left({q_s^2\over 4\pi M}\right)^2 \cos^2\theta
\end{eqnarray}  
where $\Theta$ ($\theta$) being respectively the angle between 
the observation direction ${\bf \hat R}$ and ${\bf B}^{em}$  ( 
${\bf \hat R}$ and the wave vector ${\bk}$). The crossections in 
(\ref{crossection}) is
the dual Thomson formula, while the other three also have 
dual partners in those of W bosons. As expected, the duality 
is still present
to this order including the radiations.

Notice that there exist two parameters, ${a\omega}$ 
and ${\omega/m}$ for which the above perturbative 
scheme may be improved
to their higher orders. Ignoring the terms of $O(a^2\omega^2)$  
implies that one is only interested in the linear responses 
of the BPS monopole
to the incident fields, so it is basically a weak-field 
approximation. We 
shall not improve the above treatment to this direction 
because the next
order equations are far more complicated than (\ref{perteq})-
(\ref{bperteq}), 
and more 
importantly, the most we want to pursue is already in 
the linear responses.

Proceeding to the other direction involves simply solving 
the equation (\ref{helmholtz}) to the next orders 
in $\omega/m$. As we will see, one obtains radiation damping of
the BPS monopole 
from the next order, while one finds the finite size effects 
of the monopole enters  the force law by going one-step further.
In the subsequent sections, these phenomena shall be exploited 
in detail.

\section{Radiation reaction of a BPS monopole}

As is well known in classical electrodynamics,  a motion of 
charged object subject to an external field necessarily emits 
radiations
due to an acceleration. Since the radiation carries off energy 
and momentum,
the subsequent motion of the object should be affected by the 
emission. 
Being this reaction force reducing the acceleration of the 
object, it is 
known as the phenomena of radiation damping of accelerations.  
Within the context of the
classical electrodynamics, the reaction force was analyzed 
previously  
under the following assumptions: the 
charge distribution is rigid and the  whole mass of the object 
arises
from the electromagnetic self fields\cite{abraham}. Under 
these assumptions, 
the total momentum conservation of the system leads to the 
following 
modification to the Lorentz force law; its 
nonrelativistic  form reads\cite{abraham}  
\begin{equation}
\label{dampingf}
 {\bf F}_{\rm ext} =m_{\rm cl}  {d^2\over dt^2}{\bf X}- 
{2\over 3}
{q_e^2\over 4\pi}{d^3\over dt^3}{\bf X}
 +\sum^\infty_{n=4}{(-1)^n\over 4\pi}{d^n\over dt^n}{\bf X} 
\int d{\bf r}d{\bf r'} \rho({\br})|\br-\br'|^{n-3}\rho(\br')
\end{equation}
where $q_e$ being the total charge--the spatial integration of  
electric 
charge density $\rho$.
The external force and the mass are respectively defined by 
\begin{eqnarray} 
\label{massdef}
&&{\bf F}_{\rm ext}=\int d{\bf r}(\rho{\bf E}_{\rm ext} + 
{\bf j}\times {\bf B}_{\rm ext}),\\
\label{amassdef}
&&m_{\rm cl}={1\over 2} \int d{\bf r}d{\bf r'} 
{\rho({\br})\rho(\br')\over|\br-\br'|}.
\end{eqnarray}

The second on the left side of (\ref{dampingf}) is the damping 
force, while the 
finite-size effects are described by
the terms linear in the higher time derivatives of ${\bf X}$. 
For the case of 
point charges, 
the remaining terms vanish and the radiation reaction only
depends on the total charge. Partly because of the 
nonrelativistic 
nature of the expression,
its form is linear in the time derivatives of the 
position and  the external force. As noted in 
Ref.~\cite{jackson}, 
some unsatisfactory 
features are present 
within the model. First of all, the localized distribution of 
nonvanishing total
charges requires non-electromagnetic forces holding it stable, 
and even assuming 
the existence of such forces, the charge distribution cannot 
be rigid under 
external
perturbations.  Moreover, for point charges, the mass in 
(\ref{amassdef}) 
gives rise to infinity 
that might invalidate the model explaining the realistic 
particles.

In contrast to the Abraham-Lorentz model, the radiation 
reaction effects of the
BPS monopole is well posed in the following sense. The mass of 
the 
BPS monopole is finite and the nonelectromagnetic holding 
forces are 
indeed present within the system; part of them comes from the 
attractive interaction of Higgs. The classical dynamics is 
totally governed by
the field equation (\ref{fieldeq})-(\ref{pfieldeq}), i.e. 
it is self-contained. Considering 
the radiation reaction of the BPS monopole in the presence of 
weak incident 
waves,
one need to find an appropriate solution to the equation in 
(\ref{helmholtz}).
One should note that all the linear responses of the BPS 
monopole 
to the incident waves, are included in the dynamics that is 
governed 
by the equation (\ref{helmholtz}). Since the force law 
in (\ref{lorentza}) 
from the leading order solution of Eq.(\ref{helmholtz}) does not 
include the reactive effects, one  has to solve the 
higher order terms in $O(\omega/m)$ to explore the effects.  

Due to the spherical symmetry of  the static BPS monopole, 
the presence of
incident plane waves with the wave vector $\bk$, still 
respects the axial 
symmetry
around the $\bk$ axis. The most general functional form
possessing the axial symmetry is
\begin{equation}
\label{ansatzb}
 b_{i}^a(\br)=\mp i\omega^2 a_i f [ U(r,\theta) {\hat r}^a + 
V(r,\theta){\hat\theta}^a]
\end{equation}
where $\theta$ being the angle between ${\bf\hat r}$ and 
${\bf\hat k}$. 
Inserting this 
form into the equation (\ref{helmholtz}), one obtains 
\begin{eqnarray} 
\label{basiceq}
&&\left(\nabla^2_y +{\omega^2\over m^2}\right)U -
{2\over \sinh^2 y}U -
{2\over y\sinh y}(\partial_\theta V + \cot\theta V) =0,\\
\label{abasiceq}
&&\left(\nabla^2_y +{\omega^2\over m^2}-1\right)V 
-{V\over y^2\sin^2 \theta} -
2\left({1\over \sinh^2 y}-{\coth y \over y}\right)V 
+{2\over y\sinh y}\partial_\theta U =0.
\end{eqnarray}
where one introduces a dimensionless variable, $y\equiv m r$. 
When the 
frequency $\omega$ is smaller than the W boson mass $m$, the 
equation 
(\ref{abasiceq}) tells us that V is exponentially decaying at 
large $y$. 
Using this large $y$ behavior of the function $V$, one finds 
that 
the equations for 
$U$ and $V$ are reduced to
\begin{eqnarray} 
\label{outeq}
&&\left(\nabla^2_y +{\omega^2\over m^2}\right)U =0,\ \ \ 
V=0 \ \ \ (y \gg 1)
\end{eqnarray} 
where  only the exponentially decaying terms are suppressed. 
Because of the spherical symmetry of the monopole, the 
scattering term 
of a solution of the above
equation also
possesses the symmetry to the linear responses\footnote{
 In Maxwell theory, 
one may illustrate this phenomena by considering, for example, 
the vector potential ${\bf A}$ for the case of 
spherically symmetric charge
distributions with harmonic time dependence. It is given by
${\bf A}=\int d\br' {\bf j}(\br',t)
{e^{i\omega|\br-\br'|-i\omega t}\over |\br-\br'|}$
$={\dot{\bf X}}{e^{i\omega(r-t)}\over r}\int d\br' 
\rho(r'){e^{i\omega r'}\over r'}+
O({\dot X}X)$. 
Thus, to the linear response, the scattering solution does 
respect the 
symmetry.}. 
Having 
this fact in
mind and the plane wave incidence, one concludes that the 
solution $U$ outside 
the core region of the monopole should be of the form,
\begin{equation}
\label{outsol}
 U= e^{i\bk\cdot \br} -N({\omega}){e^{i\omega r}\over mr} \ \ 
\  (y\gg 1)
\end{equation}
which may be served as an asymptotic conditions for the exact 
solution.
We expand the functions $U$ and $V$ in a power series of 
$\omega/m$ 
\begin{eqnarray} 
\label{pertuv}
&&U(r,\theta)=\sum^{\infty}_{n=0}\left({\omega\over m}\right)^n
U_{(n)}(r,\theta)\\
\label{apertuv}
&&V(r,\theta)=\sum^{\infty}_{n=0}\left({\omega\over m}\right)^n
V_{(n)}(r,\theta)
\end{eqnarray}
and 
solve the equation (\ref{basiceq})-(\ref{abasiceq}) order 
by order. Inserting the
expansion (\ref{pertuv})-(\ref{apertuv}) to (\ref{basiceq})-
(\ref{abasiceq}), 
one finds the $n^{{\rm th}}$ order 
equations ($n \ge 0$)  satisfy
\begin{eqnarray} 
\label{basiceqa}
&&\!\!\!\!\!\!\nabla^2_yU_{(n)} -{2\over \sinh^2 y}U_{(n)} -
{2\over y\sinh y}(\partial_\theta V_{(n)} + \cot\theta V_{(n)}) 
=-U_{(n-2)},\\
\label{basiceqb}
&&\!\!\!\!\!\!(\nabla^2_y-1)V_{(n)}\!-\!{V_{(n)}\over 
y^2\sin^2 \theta}  
\!-\!
2\left({1\over \sinh^2 y}\!-\!{\coth y\over y}\right)V_{(n)}\! 
+\!
{2\over y\sinh y}\partial_\theta U_{(n)} \!=\!-V_{(n-2)}
\end{eqnarray}
where $(U_{(n)}, V_{(n)})$ for $n=-2,-1$ are introduced for 
convenience and 
simply vanish.

The generic solutions in each order comprize a 
particular solution together with homogeneous
parts, whose coefficients in each order can be determined by 
comparison 
with the asymptotic form in (\ref{outsol}). In fact the zeroth 
order equation
is homogeneous and there exists a unique spherically 
symmetric nonsingular solution,
\begin{eqnarray} 
\label{zeroth}
&&U_{(0)}(r)= \coth y-{1\over y}, \ \ \ \ V_{(0)}(r)=0
\end{eqnarray}
which is  consistent with the asymptotic form (\ref{outsol}). 
For example, the other spherically symmetric solution 
$U_{(0)}={\coth y/ y}$ with vanishing $V_{(0)}$ is singular at 
the origin and 
inconsistent with the asymptotic form. This solution fixes $N$ 
to be
$1+O(\omega)$, which is in   good agreement with the previous 
analysis in 
(\ref{solutiona}). As it should be, the short distance 
behaviors of this zeroth order solution match with those in 
(\ref{solutiona}).
The $n=1$ equation 
is also homogeneous. Requiring nonsingularity at the origin and 
consistency with the asymptotic form  uniquely fix the solution 
of the 
first order equation again, and it reads
\begin{eqnarray} 
\label{first}
&&U_{(1)}= i y \cos\theta  \coth y-i\left(\coth y-
{1\over y}\right), \ \ \ 
V_{(1)}= -i {y\sin\theta\over \sinh y}.
\end{eqnarray}

The higher order solutions are readily solved by iterations once
the zeroth and the first solutions are provided.  By finding 
a particular solution and adding  
an appropriate  homogeneous solution  in 
conformity with (\ref{outsol}) in a similar fashion to
the lower order, one is led to a desired solution to the second 
order equation:  
\begin{eqnarray} 
\label{second}
&&U_{(2)}= {y\over 2}- {y^2\over 2}\coth y\cos^2\theta
-\left(\coth y-{1\over y}\right)\\
\label{asecond}
&&V_{(2)}= {y^2\sin\theta\cos\theta\over 2\sinh y}.
\end{eqnarray} 
When the relation $(\sin\theta) \,\,{\bf\hat\theta} =
-{\bf\hat k} +{\bf\hat r} ({\bf\hat\bk}\cdot{\bf\hat\br})$ is 
used, 
the solution 
$b^a_i(\br)$ to the second order in $\omega/m$ is 
summarized in the expression,
\begin{eqnarray} 
\label{solsum}
b^a_i(\br)&=& \!\mp i\omega^2 a_i f\coth y\left(1\!+
\! i\omega {\hat\bk}\cdot{\hat\br}\!+\!
{(i\omega{\hat\bk}\cdot{\hat\br})^2\over 2}\right){\hat r}_a 
\!\mp\! 
i\omega^2 a_i f {2 ir\omega\! +\!{(ir\omega)^2}{\hat\bk}\cdot
{\hat\br}
\over 2\sinh y}({\hat k}_a\!-\!
{\hat r}_a {\hat\br}\cdot {\hat\bk})\nonumber\\
& & \! \pm\! 
i\omega^2 a_i{1\over er} \left(1+i\omega r \coth y\! 
+\!{(ir\omega)^2\over 2}\right)
\left[1-{i\omega\over m}+\left({i\omega\over m}\right)^2\right]
{\hat r}_a.
\end{eqnarray}   
Consequently, the unknown function $N(\omega)$ is now obtaind 
from (\ref{solsum}) 
by the comparison of their asymptotic forms,
\begin{eqnarray} 
\label{scatt}
&&N(\omega)= 1-{i\omega\over m}+\left({i\omega\over m}\right)^2
+O({\omega^3\over m^3}).
\end{eqnarray}

Inserting the expression (\ref{solsum}) to (\ref{aperteq2}), 
using the 
ansatz in (\ref{aaansatz}), and evaluating the zero of the 
Higgs field 
$\phi_a(\br,t)$, one verifies the position is
\begin{eqnarray} 
\label{position}
&&{\bf X}= i\left\{ {\bf a}\left[1-{i\omega\over m}+
\left({i\omega\over m}\right)^2\right] -
{3\over 2}\left({i\omega\over m}\right)^2 
[{\bf a}-{\hat\bk}({\hat\bk}\cdot{\bf a})]\right\}e^{-i\omega t}
+O(\omega^3),
\end{eqnarray}   
while upon usage of the relation (\ref{defb}) and the definition 
(\ref{asymp}),
one finds the force law in terms of ${\bf a}$ 
 to be
\begin{eqnarray} 
\label{force}
&&- i\omega^2 M {\bf a}e^{-i\omega t}=
[g{\bf B}_{\rm inc}^{\rm em}+ 
q_s{\bf H}_{\rm inc}]_{\br={\bf X}}
\end{eqnarray} 
and in terms of the position,
\begin{eqnarray} 
\label{forcelaw}
[g{\bf B}_{\rm inc}^{\rm em}+ q_s{\bf H}_{\rm inc}]_{\br={\bf X}}
=M {d^2\over dt^2}{\bf X}
-{g^2\over4\pi}{d^3\over dt^3}{\bf X}+{g^2\over4\pi}
{3\over 2m}\left(
{d^4\over dt^4}{\bf X}-{\hat\bk}\,
{\hat\bk}\!\cdot\!{d^4\over dt^4}{\bf X}\right)+O(\omega^5).
\end{eqnarray} 
When only electromagnetic fields are incident upon the BPS  
monopole 
(i.e. ${\bf a}\cdot \bk =0$), the force law reads
\begin{eqnarray} 
\label{emlaw}
&& g{\bf B}_{\rm em}=M {d^2\over dt^2}{\bf X}
-{g^2\over 4\pi}{d^3\over dt^3}{\bf X}+{g^2\over 4\pi}{3\over 2m}
{d^4\over dt^4}{\bf X}+O(\omega^5),
\end{eqnarray} 
while for the  Higgs  incidence alone (i.e. ${\bf a}\times \bk 
$=0),
\begin{eqnarray} 
\label{higgslaw}
&& q_s{\bf H}=M {d^2\over dt^2}{\bf X}
-{g^2\over4\pi}{d^3\over dt^3}{\bf X}+O(\omega^5).
\end{eqnarray}

As far as the radiation dampings  in (\ref{forcelaw}), 
(\ref{emlaw}) and (\ref{higgslaw}), are concerned, they agree 
with the 
naive expectations. Namely, the radiation damping arises from 
both
the electromagnetic and the Higgs radiations. The 
electromagnetic 
part contributes to the force law by 
$-{2\over 3}{g^2\over4\pi}{d^3\over dt^3}{\bf X}$ as in 
the case of the Abraham-Lorentz model(cf. (\ref{dampingf})). 
Since the total
energy flux of the Higgs radiation is a half of the 
electromagnetic flux, 
so does its contribution to the force law. This explains the 
numerical factor
of the damping force in (\ref{forcelaw}). The next order 
terms in the force
law account for the finite size effect in the reactive forces. 
One sees clearly that the 
characteristic size relevant to the effect is none other than 
the size of
the BPS monopole. It is interesting to note that the finite 
size effect
is not present with the Higgs incidence alone, while the 
effect does exist
when the electromagnetic waves are incident upon the BPS 
monopole.
As seen in the force law (\ref{forcelaw}), the reaction 
effects---specifically 
the finite size effect---depend explicitly 
upon the direction of wave incidence, i.e. the 
wave vector ${\bf k}$, which is highly contrasted to the 
effect of 
the Abraham-Lorentz model in (\ref{lorentza})\footnote{The 
explicit 
dependence on $\bk$ does not present even considering the 
generalized Abraham-Lorentz model where the object consists 
of both electric and 
scalar charges.}. 
 Reminding that the radiations off the BPS monopole 
possess the spherical symmetry, the directional dependence 
of the reactive 
effect
is rather unusual. Presumably, the dependence is originated 
from the soft 
structures of the BPS monopole. Finally, based on the above 
results, the related
differential crossections are found to the order $\omega^2/m^2$:
\begin{eqnarray} 
\label{corsection}
&&\left({d\sigma\over d\Omega}\right)_{{\rm em}\rightarrow 
{\rm em}}=
\left({g^2\over 4\pi M}\right)^2 \left(1-{\omega^2\over m^2} 
\right) 
\sin^2\Theta,\\
&& \left({d\sigma\over d\Omega}\right)_{{\rm em}\rightarrow 
{\rm Higgs}}=
\left({g^2\over 4\pi M}\right)\left({q_s^2\over 4\pi M}\right) 
\left(1-{\omega^2\over m^2}\right) \cos^2\Theta\\
&&\left({d\sigma\over d\Omega}\right)_{{\rm Higgs}\rightarrow 
{\rm em}}=
\left({q_s^2\over 4\pi M}\right)\left({g^2\over 4\pi M}\right) 
\left(1-{\omega^2\over m^2}\right) \sin^2\theta, \\
&& \left({d\sigma\over d\Omega}\right)_{{\rm Higgs}\rightarrow 
{\rm Higgs}}=
\left({q_s^2\over 4\pi M}\right)^2 \left(1-
{\omega^2\over m^2}\right)
\cos^2\theta
\end{eqnarray}  
The corrections to the dual Thomson formula vanish 
in case of point charges, which again reflects they arises 
from the finite size
effects.

\section{W bosons and their radiation reactions}

Though the spin contents do not match with each other, the dual 
partners of the BPS monopoles in the theory are known to be 
the W bosons.
In the $N=4$ supersymmetric Yang-Mills theory, even their  
spin contents 
exactly match\cite{seiberg}.
However, we shall adhere to our present theory since it 
possesses  
all the essential 
ingredients for our purpose. In this section, 
we shall describe how W bosons couple to
the electromagnetic fields and the Higgs fields. As we will 
see below 
explicitly, 
the couplings are dual to those of the BPS monopole. 
Namely, W bosons couple to photons and Higgs with coupling 
strengths 
$e$ and $e$ 
respectively. On the other hand, for the BPS monopole, the 
coupling
constants of the dual photons and Higgs are respectively 
$g$ and $q_s$ $(=g)$ 
with the condition  ${eg/ 4\pi}= 1$. The masses are again 
dual in the sense that 
the W boson mass $ef$ is obtained from  the monopole mass 
$gf$ by 
interchanging the magnetic 
charge with the electric charge.  To compare the dynamics of 
the W boson to
those of the BPS monopole in the last section, we shall  first 
describe
nonrelativistic dynamics of W bosons , which will be obtained 
from a systematic
nonrelativistic reduction of the Lagrangian (\ref{lag}). The 
validity of 
the nonrelativistic version is limited by the condition that 
the velocity
of the W boson should be much smaller than the light velocity. 
Upon 
consideration of
dynamical processes whose leading order is $O(v)$, one finds that
the nonrelativistic approximation 
corresponds to ignoring $O(v^2)$ terms, which in turn 
corresponds to the 
linear-response approximation of the previous sections.

To begin, let us choose  the unitary
gauge  where one may put $\phi^a=(0,0,f+\varphi(\br,t))$ with a 
real scalar field $\varphi(\br,t)$. We may rewrite the Lagrange 
density (\ref{lag}) as
\begin{eqnarray}
\label{laguni}
&&{\cal L}=-{1\over 4} F^{\mu\nu}F_{\mu\nu}-{1\over 2}
(\dd_\mu W_\nu-\dd_\nu W_\mu)^\dagger 
( \dd^\mu W^\nu-\dd^\nu W^\mu) 
-{1\over 2} \partial_\mu \varphi \partial^\mu \varphi 
-m^2 c^2 {W^\mu}^\dagger W_\mu\nonumber\\
&&\ \ \ +\!{4e^2\over c^2}
({W_\mu}^\dagger W_\nu\!-\!{W_\nu}^\dagger W_\mu) 
( {W^\mu}^\dagger W^\nu\!-\!
{W^\nu}^\dagger W^\mu)
\!-\!2e m \varphi  {W^\mu}^\dagger W_\mu 
 \!-\!{e^2\over c^2} \varphi^2 {W^\mu}^\dagger W_\mu 
\end{eqnarray}
where we rename the gauge fields with
\begin{eqnarray}
&&A_{\mu}^{\rm em} = A^{(3)}_\mu, \ \ \ \ 
 W_\mu= {1\over \sqrt{2}} (A^{(1)}_\mu -i A^{(2)}_\mu),
\end{eqnarray}
and define the covariant derivative as 
\begin{eqnarray}
\dd_\mu W_\nu\equiv (\partial_\mu +i{e\over c} A_{\mu}) W_\nu.
\end{eqnarray}
Here we recover the light velocity $c$  in order to find 
the nonrelativistic limit.

Note that the field equation for the W boson
reads
\begin{eqnarray} 
\label{wequation}
&&(\dd_\nu \dd^\nu -m^2 c^2)W^\mu -
\dd^\mu \dd^\nu W_\nu\nonumber\\
&&=\left[{2ie\over c}F^{\mu\nu}\!+\!{e^2\over c^2}
( {W^\mu}^\dagger W^\nu\!-\!
{W^\nu}^\dagger W^\mu)\right]W_\nu\! +\!
\left(2em \varphi\! -\!{e^2\over c^2}\varphi^2\right)W^\mu.
\end{eqnarray}    
Owing to the fact that the zeroth component of 
the above equation does not involve time derivatives, it is  
a constraint 
equation. 
We solve this constraint by expressing $W_0$ in terms 
of the others. 
In the nonrelativistic limit, one may get an explicit 
expression for $W_0$,
in the $c\rightarrow \infty$ limit, as
\begin{eqnarray} 
\label{wlimit}
&&W_0={i \over m c \sqrt{2m}}e^{-im c^2 t} \dd_i\psi_i +O(c^{-2})
\end{eqnarray}     
where the field $\psi_i$ is defined by the relation 
$W_i ={e^{-im c^2 t}\over \sqrt{2m}} \psi_i$ omitting 
anti-particle sector.

In terms of these variables, one finds the following 
 desired nonrelativistic expression for action  (\ref{laguni}):
\begin{eqnarray}
\label{lagnon}
{\cal L}\!=\!-{1\over 4} F^{\mu\nu}F_{\mu\nu}
-{1\over 2} \partial_\mu \varphi \partial^\mu \varphi  
+i\psi_i^\dagger \dd_t \psi_i -
{1\over 2 m} (\dd_i \psi_j)^\dagger \dd_i \psi_j 
-e\varphi \psi_i^\dagger \psi_i -{ie\over m c} B_k 
\epsilon^{kij}\psi_i^\dagger \psi_j
\end{eqnarray}
Here, it is clear that there are three spin degrees 
of freedom for the W boson and they couple to
the  photon as well as  the massless 
Higgs with coupling strength $(e,\,\,\,e)$. Because of the 
internal structure of the W boson, we have the spin-one Pauli 
term meaning that the W boson
possesses a nonvanishing magnetic moment.

The above Lagrangian may further be 
reduced to a particle Lagrangian;
\begin{eqnarray}
\label{lagpar}
&&{L}\!={1\over 2}m {\bf\dot X}\!\cdot\!{\bf \dot X}\!+
\!e A_0^{\rm em}\!
+\!e\varphi\!-\!
{e\over c}{\bf\dot X}\!\cdot\! {\bf A^{\rm em}} \!+
\!i{\rm tr}(K g^{-1}  {\dot g})\! - \!{e\over m c} B_k I_k 
\nonumber\\
&&\ \ \ \ \ \ \ -
\int \! d\br\left({1\over 4} F^{\mu\nu}F_{\mu\nu}
\!+\!{1\over 2} \partial_\mu 
\varphi \partial^\mu \varphi\right)  
\end{eqnarray}
where $g(t)$ is the $SO(3)$ group element and the constant 
element 
$K$ belongs to the corresponding Lie algebra.
The term in the trace is the Kirillov-Kostant one-form 
providing symplectic 
structure
for the classical spin variable  $I_k(t)$ that is defined by
\begin{eqnarray}
\label{cspin}
I_k(t) T_k \equiv g(t) K g^{-1}(t) 
\end{eqnarray}
where $T_k$ are the three generators for the $SO(3)$ 
group\cite{bak3}. Utilizing 
the symplectic structure, 
it can shown that 
the Poisson braket for the spin variable is 
$\{I_i,I_j\}=\epsilon_{ijk}I_k$. Thus, 
upon quantization,  
they are realized by  matrices satisfying 
$[I_i,I_j]=i\epsilon_{ijk}I_k$. 
However, for our present purpose, this digression is 
not necessary because
we shall ignore the magnetic moment interaction below. 
As said earlier, as far as spins are concerned, 
the duality is already explicitly broken in our model. 
Since this mismatch
disappears in the $N=4$ supersymmetric theory, we shall 
not further pursue
its consequence below.     

 Though there are  Higgs 
radiations of the W particle in addition to the 
electromagnetic ones, 
 the framework of the Abraham-Lorentz
model can be generalized  to compute the contribution of 
the radiation reaction 
to the force law. A straightforward analysis leads to a 
force law:
\begin{eqnarray} 
\label{wforce}
&& [e{\bf E}_{\rm inc}^{\rm em}+ 
e{\bf H}_{\rm inc}]_{\br={\bf X}}
=m {d^2\over dt^2}{\bf X}
-{e^2\over4\pi}{d^3\over dt^3}{\bf X}.
\end{eqnarray}    
The corresponding scattering crossections of the W boson to 
(\ref{crossection})-(\ref{ccrossection}) are 
given as the dual forms of 
(\ref{crossection})-(\ref{ccrossection}), 
i.e. the ($g$,$\,q_s$) are 
replaced with ($e$,$\,e$), $M$ with $m$, and the  $\Theta$ angle 
becomes the one between
the electric field and the observation direction. 

The direct comparison of the force law of the monopole in 
(\ref{forcelaw}) to 
that of W boson in (\ref{wforce}) is finally at hand. The 
damping of accelerations 
are dual symmetric, while the finite size effects to the 
reaction, are present only in 
the case of the BPS monopole. In the crossections, the dual 
symmetry is also broken 
 by the size effects. 

A few comments are in order. First, let us 
observe the fact that the quantization of the modular space 
parameters ${\bf X}$, 
inevitably introduces the Compton size of the BPS monopole, 
which is the inverse of the monopole mass, i.e. $1/M$. This 
scale may be 
explicitly seen in the Compton scattering of the BPS monopole 
that is basically a {\it photon} scattering off the BPS 
monopole. Therefore, 
one finds 
two scale parameters  are present for  dynamical processes 
of the BPS monopole.
Namely, one is the Compton size of the monopole $1/M\,\,
[=1/(gf)]$ and 
the other is the classical size  $1/m\,\,[=g/(4\pi f)]$. 
When the coupling between monopole and 
photon is weak ( i.e. $g \ll 1$), the classical size is 
smaller than 
the Compton size. Nevertheless, the effect of the classical 
size cannot be
ignored.

Considering the Compton scattering of the W boson 
also involves its Compton size, $1/m$. For example,  
the Compton scattering cross-section that is computed from 
the Lagrangian 
(\ref{lag}) to the tree level depends upon the scattered 
photon energy 
$\omega'$
\begin{eqnarray} 
\label{compton}
&& \omega'=\omega \left(1+{\omega\over m}(1-\cos\theta)
\right)^{-1}
\end{eqnarray}     
where $\omega$ being the energy of the incident photon. Since 
the classical descriptions of the monopole 
and W-boson are dual to each other except
the finite size nature of the monopole, and the effects of 
Compton size 
are entering the problem as a
consequence of quantizations, one may argue that the Compton 
scattering  of the W 
boson is dual to that of the monopole ignoring the 
finite size effect of the 
monopole.  
At least, a pair of dual scales---the Compton sizes of 
the monopole and W boson---is present within the theory.

On the other hand, for the 
case of finite size effects, naively there seems to be 
no scale dual to the 
size since the W boson is pointlike. This explains 
why the dual symmetry is
broken in the force laws  (\ref{forcelaw}) and 
(\ref{wforce})\footnote{
The physical situation here is different from
the weak and strong coupling duality between sine-Gordon 
solitons and
elementary excitations in massive Thirring model. 
Although the sine-Gordon 
solitons have a finite size and the Thirring excitations are 
elementary (pointlike), the mapping is exact for all 
the dynamical processes. 
This is possible because  there is no finer excitations 
to probe their
structures within the sine-Gordon model. On the contrary, 
in the case 
of the BPS monopole
the photon can be used to probe its structures.}.  The 
dual symmetry might be
saved if one includes contributions from virtual 
creation of monopole--antimonopole pairs around W bosons. 
Since the 
energy scale involved in the process
is the monopole mass,  the length scale appears to be $1/M$ 
through $\omega/M$. Though not clear, this size
of the virtual clouds might be dual to the size of the 
BPS monopole\footnote{
  Even included the effect of monopole-antimonopole virtual 
creation, 
the monopole being neutral in their electric charges, still the 
W boson seems pointlike
in their charge distribution.}. Assuming the dual symmetry 
holds even for the finite size effect,
one finds there should be a corresponding dual size effect 
for the motion of a W boson. This effect is 
completely missing in the force law (\ref{wforce}), while 
the above 
consideration 
suggests that the virtual
pair creation  
be the candidate for the process. Hence, upon the assumption 
of the dual 
symmetry, the contributions from the  pair creation are obtained 
by the duality transformation from the result of the BPS 
monopole (e.g. (\ref{forcelaw})). 
However it should be commented that it is never clear the 
process of pair creation  is really dual 
to the finite size effects of
the BPS monopole, so is the dual symmetry on the finite 
size effect. 

Another interesting
feature of the Lagrangian (\ref{lag}) is that it  also 
possesses dyonic 
solutions\cite{julia}. 
Notice that the dyons have a  
 size proportional to ${\sqrt{g^2+q^2}}/ (4\pi f)$, where 
$g$ and $q$ 
denote respectively 
its magnetic and  electric charges. In Ref.~\cite{bak2}, it 
is shown that 
the dyons also emit radiations
when they are accelerated. (The equation similar to Eq. 
(\ref{perteq}) that 
describes the response of a dyon to
the incident waves, is derived in Ref.~\cite{bak4} from 
different context.) 
As a consequence, one expects 
that the damping and the 
finite size effects also exist for the dyons. Its detailed 
description and 
implication on the duality
needs further investigations.

\acknowledgements

It is a pleasure to thank B.H. Lee, C. Lee and K. Lee for 
enlightening discussions.
This work was
supported in part by funds provided by 
the Korea Science and Engineering Foundation through
the SRC program of SNU-CTP, and
the Basic Science Research Program
under projects \#BSRI-96-2425 \#BSRI-96-2441.

\end{document}